\documentclass[12pt]{article}

\usepackage{amssymb}
\usepackage{amsmath}
\usepackage{latexsym}

\begin{document}

\textheight=24cm
\textwidth=16.5cm
\topmargin=-1.5cm
\oddsidemargin=-0.25cm.




\renewcommand{\vec}[1]{{\bf #1}}       
\def\beq{\begin{eqnarray}}    
\def\eeq{\end{eqnarray}}      

\def\ln{\,\mbox{ln}\,}                  
\def\tr{\,\mbox{tr}\,}                  
\def\Tr{\,\mbox{Tr}\,}                  
\def\Box{\square}                       
\def\det{\,\mbox{det}\,}                
\def\Det{\,\mbox{Det}\,}                
\def\Res{\,\mbox{Res}\,}                
\def\sTr{\,\mbox{sTr}\,}                
\def\str{\,\mbox{str}\,}                
\def\sDet{\,\mbox{sDet}\,}              
\def\Ber{\,\mbox{Ber}\,}                
\def\fat{\;\mbox{для любого}\;}

\renewcommand{\Re}{\,\mbox{Re}\,}       
\renewcommand{\Im}{\,\mbox{Im}\,}       

\begin{center}

{\Large\bf Superfield generating equation of field-antifield
formalism }

\vspace{8mm}

{\large I. A. Batalin$^{a}$
and P. M. Lavrov$^{b}$
}


${{}^{a}}$
{\em P.N. Lebedev Physics Institute,
Leninsky Prospect \ 53, 119991 Moscow, Russia}

${{}^{b}} ${\em
Tomsk State Pedagogical University,
Kievskaya St.\ 60, 634061 Tomsk, Russia}

E-mails: batalin@lpi.ru, lavrov@tspu.edu.ru


\vspace{5mm}

\begin{abstract}
\noindent
A  simple quantum superfield  generating  equation of  the field-antifield  formalism
 is proposed. The  Schroedinger  equation with the  Hamiltonian having  $\Delta$-exact
form is derived. An $Sp(2)$ symmetric extension
to the main construction, with specific features caused by the principal fact
that all basic equations become $Sp(2)$ vector-valued ones, is presented. A principal role
of quantum antibrackets in formulation of the Heisenberg equations of motion is shown.
\end{abstract}

\end{center}

\section{Introduction}
The field-antifield formalism or Batalin-Vilkovisky method is a
unique closed approach to the problem of covariant quantization of
general gauge theories \cite{BV,BV1}. The quantum master equation
formulated in terms of the odd nilpotent Laplacian operator ($\Delta
$-operator) for a quantum action plays a fundamental role in
deriving  all basic properties of the formalism (gauge independence
of $S$-matrix, Ward identity, gauge invariant renormalization and so
on).

In the present paper we present a new sight on the field-antifield
formalism at the quantum superfield level. Firstly, within a
superfield approach \cite{BBD1,BBD2,BL-IJMP}, we formulate a simple
quantum generating equation of the field-antifield formalism as
having its configuration space identified with the antisymplectic
phase space of fields and antifields. The latter generating equation
is presented in terms of a superfield covariant derivative with
respect to the two-dimensional super-time whose Boson component is
the "ordinary" time, purely formal in its origin, while its Fermion
component is identified naturally with the BRST parameter. The
covariant derivative squared is just the "ordinary" time derivative.
Then we derive the standard Schroedinger equation by applying again
the covariant derivative to the  generating superfield equation. We
provide effectively for the Hamiltonian commuting with the
$\Delta$-operator and being  a $\Delta$-exact one. Choosing a
special form of the wave function in this Schroedinger equation we
reproduce the quantum master equation of the field-antifield
formalism. We present an $Sp(2)$ symmetric extension  to the main
construction, with specific features caused by the principal fact
that all basic equations become $Sp(2)$ vector-valued ones.

\section{Phase space}
Our starting point is the phase space for which the co-ordinate operators $Z^{ A }$,
\beq
\label{SBV1}
Z^{A}  =:  ( \Phi^{\alpha} ; \Phi^*_{\alpha} ), \quad \varepsilon( \Phi^{ \alpha}
)  =  \varepsilon( \Phi^*_{ \alpha } )  +  1,     
\eeq
are identified with the standard full set of the field-antifield variables, and $P_{ A }$,
\beq
\label{SBV2}
P_{A}  =:  - i \hbar \partial_{A}  (-1)^{ \varepsilon_{A} }\quad  \Rightarrow \quad [ Z^{A},
P_{ B} ]  =  i \hbar \delta^{A}_{B}.    
\eeq
are their respective canonically-conjugate momenta operators.

\section{ Superfield }
We introduce a superfield  $\Psi$,
\beq
\label{SBV3}
\Psi  =: \Psi( Z; t, \tau ),\quad  \varepsilon( t )  =  0,  \quad\varepsilon( \tau )
=  1.    
\eeq
as a function depending on the co-ordinates of phase space and
the "time" $t$ and its superpartner $\tau$.

\section{ Superfield dynamics }
We note that the generating equation of the
field-antifield formalism takes the very simple form of a superfield
Schroedinger equation,
\beq
\label{SBV4}
( i \hbar D  -  Q ) \Psi  =  0,   
\eeq
where $D$ is a covariant super-time derivative,
\beq
\label{SBV5}
D  =:  \partial_{ \tau }  +  \tau \partial_{ t },\quad  \varepsilon( D )  =  1,\quad
[ D, D ]  =  2 \partial_{ t },   
\eeq
$Q$ is a super-charge
\beq
\label{SBV6}
Q  =:  \Delta  -  F, \quad \varepsilon( Q )  =  \varepsilon( \Delta )  =
\varepsilon( F )  =  1,   
\eeq
whose kinetic part is the odd Laplacian, $\Delta$,
\beq
\label{SBV7}
\Delta  =:  \frac{1}{2}  P_{A} E^{AB} P_{B}  (-1)^{ \varepsilon_{ B} },\quad
E^{AB}  =  {\rm const},  
\eeq
with $E^{AB}$ being antisymplectic structure,
\beq
\label{SBV8}
\varepsilon( E^{AB} )  =:  \varepsilon_{A}  +  \varepsilon_{B}  +  1,  
\eeq
\beq
\label{SBV9}
E^{AB}  =  -  E^{BA}  (-1)^{ ( \varepsilon_{A} + 1 ) ( \varepsilon_{B} + 1 )
}, \quad [ \Delta, \Delta ]  =  0,   
\eeq
and $F$ is a super-potential,
\beq
\label{SBV10}
F  =:  F( Z )\quad  \Rightarrow \quad [ F, F ]  =  0,
\eeq
It follows from (\ref{SBV4}) that the standard Schroedinger equation holds,
\beq
\label{SBV10a}
( i \hbar \partial_{ t }  -  H ) \Psi_{0}  =  0,
\eeq
with the Hamiltonian $H$,
\beq
\label{SBV11}
H  =:  -  \frac{1}{2} (i \hbar )^{-1} [ Q, Q ]  = ( i \hbar )^{-1}  [ \Delta, F ],
\eeq
and the decomposition of superfield $\Psi$,
\beq
\label{SBV11a}
\Psi(Z;t,\tau)  =  \exp\{ \tau ( i \hbar )^{-1} Q \}\Psi_{0}( Z; t ).
\eeq
Due to the nilpotency of $\Delta$-operator and the definition (\ref{SBV11}), we
have
\beq
\label{SBV11b}
[ \Delta,  H ]  =  0.
\eeq
From (\ref{SBV11b}) we conclude that if $\Psi_{0}$ satisfies the
Schroedinger equation (\ref{SBV10a}) then $\Delta \Psi_{0}$ is a solution to this
equation as well,
\beq
\label{SBV11c}
( i\hbar \partial_{ t }  -  H  ) \Delta \Psi_{0}  =  0.
\eeq
In its turn, it implies the following statement
\beq
\label{SBV11d}
  \Delta \Psi_{0}|_{ t = 0 }  =  0  \quad\Rightarrow\quad
  \Delta \Psi_{0}|_{ any\; t }  = 0.      
\eeq
Then, by choosing $\Psi_{0}(Z;t)=\exp\{t\frac{i}{\hbar}W(Z)\}$, we
arrive at the quantum master equation of the field-antifield formalism
\cite{BV,BV1} for quantum action $W=W(Z)$, when $t=1$,
\beq
\label{SBV11e}
 \Delta\exp\Big\{\frac{i}{\hbar}W\Big\}=0.
\eeq

\section{Sp(2) symmetric version}

In the case of $Sp(2)$-covariant quantization scheme \cite{BLT1,BLT2,BLT3,BBL}
co-ordinate part $Z^A$ of the phase space,
\beq
\label{SBV12}
Z^{A}  =:  ( \Phi^{ \alpha }, \Phi^{ \alpha a } ;  \Phi^*_{ \alpha a },
\Phi^{**}_{ \alpha }  ),     
\eeq
are identified with the set of all field and antifield variables.
The generating equation of $Sp(2)$ formalism takes the form of
$Sp(2)$ vector valued superfield Schroedinger equation 
\beq
\label{SBV13}
( i \hbar D^{a}  -  Q^{a} ) \Psi  =  0, \quad  D^{a}  =:  \partial_{ \tau_{a} }
+  g^{ab} \tau_{b} \partial_{ t },    
\eeq
where the following conventions hold for the required $Sp(2)$
vector valued operators
\beq
\label{SBV14}
[ D^{a}, D^{b} ]  =  2 g^{ab} \partial_{ t }, \quad
g^{ab}  =  g^{ba}  =  {\rm const},\quad
Q^{a}  =:  \Delta^{a}_{+}  -  F^{a},    
\eeq
\beq
\label{SBV15}
\Delta^{a}_{+}  &=:&  \Delta^{a}  +  \frac{i}{\hbar}  V^{a}, \\
\label{SBV16}
\Delta^{a}  &=:&  \frac{1}{2} P_{A} E^{ABa} P_{B}  (-1)^{ \varepsilon_{B} },\\
\label{SBV17}
V^{a}  &=:&  -  i \hbar  \varepsilon^{ab}  \Phi^*_{ \alpha b}  P^{ \alpha
}_{**}  (-1)^{ \varepsilon_{ \alpha } },  \\   
\label{SBV18}
F^{a}  &=:&  g^{ab}  \varepsilon_{bc}  ( i \hbar )^{-1}  [ \Delta^{c}_{+}, B
].      
\eeq
Choosing $B$ as a Boson functional depending on fields $\{\Phi^{ \alpha }\}$
only,
\beq
\label{SBV19}
B=B(\Phi),
\eeq
we arrive at the implication
\beq
\label{SBV19a}
  ( B,B)^{a} =  0 \quad\Rightarrow\quad[ F^{a}, F^{b} ]  =  0,   
\eeq
where the notation $(G,H)^a$ for extended antibrackets is used \cite{BM3}.
Using the component form of superfield $\Psi=\Psi(Z;t,\tau_a)$
\beq
\label{SBV20}
\Psi  =  \exp\{ \tau_{a} ( i \hbar )^{-1}
Q^{a} \}  \Psi_{0}( Z; t ),
\eeq
we find that $\Psi_{0}$ satisfies the Schroedinger equation
\beq
\label{SBV20a}
 ( i
\hbar \partial_{ t }  -  H ) \Psi_{0}  =  0,    
\eeq
with the Hamiltonian $H$,
\beq
\label{SBV21}
H  =:  -  \frac{1}{4}  g_{ab}  ( i \hbar
)^{-1}  [ Q^{a}, Q^{b} ]  = \frac{1}{2}  ( i \hbar )^{-2}  [
\Delta^{a}_{+},  \varepsilon_{ab}
[ \Delta^{b}_{+}, B ]],    
\eeq
where $g_{ab}$ is inverse to $g^{ab}$.
Due to the anti-commutativity of the operators $\Delta^{a}_{+}$,
\beq
\label{SBV21a}
[\Delta^{a}_{+},\Delta^{b}_{+}]=0,
\eeq
the Hamiltonian $H$ (\ref{SBV21}) commutes with $\Delta^{a}_{+}$,
\beq
\label{SBV22}
[ \Delta^a_{+},  H ]  =  0,
\eeq
and, as a consequence, the $\Delta^a_{+} \Psi_{0}$ satisfies the
Schroedinger equation,
\beq
\label{SBV23}
( i
\hbar \partial_{ t }  -  H  ) \Delta^a_{+} \Psi_{0}  =  0,
\eeq
if the $\Psi_{0}$ does satisfy. It follows from (\ref{SBV23}) that the implication
holds of
\beq
\label{SBV24}
  \Delta^a_{+} \Psi_{0}|_{ t = 0 }  =  0  \quad\Rightarrow\quad
  \Delta^a_{+} \Psi_{0}|_{ any\; t }  =0.      
\eeq
By choosing $\Psi_{0}(Z;t)=\exp\{t\frac{i}{\hbar}W(Z)\}$, we arrive
at the quantum master equation of the Sp(2)-covariant quantization
\cite{BLT1,BLT2,BLT3},
\beq
\label{SBV25}
  \Delta^a_{+}\exp\Big\{\frac{i}{\hbar}W\Big\}=0,
\eeq
as written for the quantum action $W=W(Z)$, when $t=1$.

\section{ Heisenberg equations of motion}
In this Section, we are going to emphasize  a principal role of the quantum antibrackets
\cite{BM1,BM2}
in formulation of the Heisenberg equations of motion
 both  in the $Sp( 1 )$ \cite{BV,BV1} and the $Sp( 2 )$ \cite{BLT1,BLT2,BLT3} cases.

Denote with $\Gamma$ the full set of the Schroedinger canonical variable  operators,
\beq
\label{SFBV70}
\Gamma  =:  (  Z^{ A } ;  P_{ A }  ),    
\eeq
and let $\tilde{ \Gamma }( t , \tau )$  be the respective superfield Heisenberg
canonical variable operators.
\beq
\tilde{ \Gamma }=\tilde{ \Gamma }( t , \tau ).
\eeq

In the $Sp( 1 )$ case, the superfield Heisenberg equations of motion have the form,
\beq
\label{SFBV71}
i \hbar  D  \tilde{ \Gamma }  =  [\tilde{ Q },  \tilde{\Gamma}] , \quad 
i \hbar  D  \tilde{ Q }  =  [  \tilde{ Q },  \tilde{ Q }  ].    
\eeq
It follows from these equations \cite{BM2},
\beq
\label{SFBV73}
( i \hbar )^{ 2 }  \frac{ \partial }{ \partial t }  \tilde{ \Gamma }  =
 -  \frac{ 1 }{ 2 }  [  \tilde{ \Gamma },  [  \tilde{ Q },  \tilde{ Q }  ]  ]  =
 -  \frac{ 2 }{ 3 }  (  \tilde{ \Gamma },  \tilde{ Q }  )_{ \tilde{ Q } },    
\eeq
where the quantum 2 - antibracket, $(A,  B)_{ Q }$, is
defined by
\beq
\label{SFBV33}
(A, B )_{Q }  =:  \frac{1}{2}  (   [  A,  [Q , B  ]  ]  -
( A \;\leftrightarrow \;  B  )
(-1)^{  ( \varepsilon_{ A }  + 1 ) ( \varepsilon_{ B } + 1 )  }  ).    
\eeq

In the $Sp( 2 )$ case,  the respective superfield Heisenberg
equations of motion have the form,
\beq
\label{SFBV74}
i \hbar  D^{ a }  \tilde{ \Gamma }  =  [  \tilde{ Q }^{ a },  \tilde{ \Gamma }  ], \quad 
i \hbar  D^{ a }  \tilde{ Q }^{ b }  =  [  \tilde{ Q }^{ a },  \tilde{ Q }^{ b }  ].   
\eeq
It follows from these equations,
\beq
\label{SFBV76}
( i \hbar )^{ 2 }  \frac{ \partial }{ \partial t }  \tilde{ \Gamma }  =
 -  \frac{ 1 }{ 4 }  g_{ ab }  [  \tilde{ \Gamma },
[  \tilde{ Q }^{ b }, \tilde{ Q }^{ a }  ]  ]  =
 -  \frac{ 1 }{ 3 }  g_{ ab }
(  \tilde{ \Gamma },  \tilde{ Q }^{ b }  )^{ a }_{ \tilde{ Q } },  
\eeq
where the $Sp( 2 )$ vector valued  quantum 2-antibracket,
$(  A,  B  )^{ a }_{ Q }$ \cite{BM4},
is defined
by
\beq
\label{SFBV50}
(  A, B  )^{ a }_{Q }   =:
\frac{ 1 }{ 2 }  \left(  [  A,  [Q^a,  B  ]  ]  -
(  A \;\leftrightarrow\; B)
(-1)^{  ( \varepsilon_{ A } + 1 ) ( \varepsilon_{ B } + 1 )  }\right),  
\eeq


\begin {thebibliography}{99}
\addtolength{\itemsep}{-8pt}

\bibitem{BV}
Batalin I. A., Vilkovisky G. A. Gauge algebra and quantization//
Phys. Lett. B. 1981. V. 102. P. 27-31.

\bibitem{BV1}
Batalin I. A., Vilkovisky G. A. Quantization of gauge
theories with linearly dependent generators// Phys. Rev. D. 1983. V.
28. P. 2567-2582.

\bibitem{BBD1}
Batalin I. A., Bering K., Damgaard P. H. Superfield quantization//
Nucl. Phys. B. 1998. V. 515. P. 455-487.

\bibitem{BBD2}
Batalin I. A., Bering K., Damgaard P. H. Superfield formulation of
the phase path integral//
 Phys. Lett. B. 1999. V. 446. P. 175-178.

\bibitem{BL-IJMP}
Batalin I. A., Lavrov P. M. Superfield Hamiltonian quantization in
terms of quantum antibrackets// Int. J. Mod. Phys. A. 2016. V. 31.
P. 1650054-1-14.

\bibitem{BLT1}
Batalin I. A., Lavrov P. M., Tyutin I. V. Covariant quantization of
gauge theories in the framework of extended BRST symmetry// J. Math.
Phys. 1990. V. 31. P. 1487-1493.

\bibitem{BLT2}
Batalin I. A., Lavrov P. M., Tyutin I. V. An Sp(2) covariant
quantization of gauge theories with linearly dependent generators//
J. Math. Phys. 1991. V. 32. P. 532-539.

\bibitem{BLT3}
Batalin I. A., Lavrov P. M., Tyutin I. V. Remarks on the Sp(2)
covariant Lagrangian quantization of gauge theories// J. Math. Phys.
1991. V. 32. P. 2513-2521.

\bibitem{BBL}
Batalin I. A., Bering K., Lavrov P. M. A systematic study of finite
BRST-BV transformations within $W-X$ formulation of the standard and
the $Sp(2)$-extended field-antifield formalism// Eur. Phys. J. C.
2016. V. 76. P. 101-1-8.

\bibitem{BM3}
Batalin I., Marnelius R. Completely anticanonical form of Sp(2)
symmetric Lagrangian quantization// Phys. Lett. B. 1995. V. 350. P.
44-48.

\bibitem{BM1}
Batalin I., Marnelius R. Quantum antibrackets// Phys. Lett. B, 1998.
V. 434. P. 312-320.

\bibitem{BM2}
Batalin I., Marnelius R. General quantum antibrackets//
 Theor. Math. Phys. 1999. V. 120. P. 1115-1132.

\bibitem{BM4}
Batalin I., Marnelius R. Quantum Sp(2) antibrackets and open
groups// Nucl. Phys. B. 1999. V. 551. P. 450-466.

\end{thebibliography}
\end{document}